\documentclass[11pt]{article}
\usepackage{amsmath,amssymb}
\usepackage{array}
\newcount\figureno     \figureno=0
\newdimen\figdim       \figdim=70mm
\def\figureinc{%
   \global\advance\figureno by 1%
}
\def\figcaption#1#2#3{\hbox to #2{\hss{\vbox{\hsize=#2 \parindent=0pt
        {\bf Figure \number\figureno#3 :\ }#1}}\hss}
}

\begin{document}
\baselineskip 100pt
\renewcommand{\baselinestretch}{1.5}
\renewcommand{\arraystretch}{0.666666666}
{\large
\parskip.2in
\numberwithin{equation}{section}
\newcommand{\be}{\begin{equation}}
\newcommand{\ee}{\end{equation}}
\newcommand{\ben}{\begin{equation*}}
\newcommand{\een}{\end{equation*}}
\newcommand{\eqalinb}{\begin{eqnarray}}
\newcommand{\eqaline}{\end{eqnarray}}
\newcommand{\br}{\bar}
\newcommand{\fr}{\frac}
\newcommand{\lm}{\lambda}
\newcommand{\ra}{\rightarrow}
\newcommand{\al}{\alpha}
\newcommand{\bt}{\beta}
\newcommand{\z}{\zeta}
\newcommand{\pa}{\partial}
\newcommand{\hs}{\hspace{5mm}}
\newcommand{\up}{\upsilon}
\newcommand{\bigb}{\hspace{7mm}}
\newcommand{\dg}{\dagger}
\newcommand{\vphi}{\vec{\varphi}}
\newcommand{\ve}{\varepsilon}
\newcommand{\acc}{\\[3mm]}
\newcommand{\dl}{\delta}
\newcommand{\sdil}{\ensuremath{\rlap{\raisebox{.15ex}{$\mskip
6.5mu\scriptstyle+ $}}\subset}}
\newcommand{\sdir}{\ensuremath{\rlap{\raisebox{.15ex}{$\mskip
6.5mu\scriptstyle+ $}}\supset}}
\def\tablecap#1{\vskip 3mm \centerline{#1}\vskip 5mm}
\def\p#1{\partial_#1}
\newcommand{\pd}[2]{\frac{\partial #1}{\partial #2}}
\newcommand{\pdn}[3]{\frac{\partial #1^{#3}}{\partial #2^{#3}}}
\def\DP#1#2{D_{#1}\varphi^{#2}}
\def\dP#1#2{\partial_{#1}\varphi^{#2}}
\def\xh{\hat x}
\newcommand{\Ref}[1]{(\ref{#1})}
\def\ld{\,\ldots\,}

\def\C{{\mathbb C}}
\def\Z{{\mathbb Z}}
\def\R{{\mathbb R}}
\def\mod#1{ \vert #1 \vert }
\def\chapter#1{\hbox{Introduction.}}
\def\Sin{\hbox{sin}}
\def\Cos{\hbox{cos}}
\def\Exp{\hbox{exp}}
\def\Ln{\hbox{ln}}
\def\Tan{\hbox{tan}}
\def\Cot{\hbox{cot}}
\def\Sinh{\hbox{sinh}}
\def\Cosh{\hbox{cosh}}
\def\Tanh{\hbox{tanh}}
\def\Asin{\hbox{asin}}
\def\Acos{\hbox{acos}}
\def\Atan{\hbox{atan}}
\def\Asinh{\hbox{asinh}}
\def\Acosh{\hbox{acosh}}
\def\Atanh{\hbox{atanh}}
\def\frac#1#2{{\textstyle{#1\over #2}}}

\def\ph{\varphi_{m,n}}
\def\phl{\varphi_{m-1,n}}
\def\phr{\varphi_{m+1,n}}
\def\varphil{\varphi_{m-1,n}}
\def\varphir{\varphi_{m+1,n}}
\def\varphit{\varphi_{m,n+1}}
\def\varphib{\varphi_{m,n-1}}
\def\pht{\varphi_{m,n+1}}
\def\phb{\varphi_{m,n-1}}
\def\phbl{\varphi_{m-1,n-1}}
\def\phbr{\varphi_{m+1,n-1}}
\def\phtl{\varphi_{m-1,n+1}}
\def\phtr{\varphi_{m+1,n+1}}
\def\u{u_{m,n}}
\def\ul{u_{m-1,n}}
\def\ur{u_{m+1,n}}
\def\ut{u_{m,n+1}}
\def\ub{u_{m,n-1}}
\def\utr{u_{m+1,n+1}}
\def\ubl{u_{m-1,n-1}}
\def\utl{u_{m-1,n+1}}
\def\ubr{u_{m+1,n-1}}
\def\v{v_{m,n}}
\def\vl{v_{m-1,n}}
\def\vr{v_{m+1,n}}
\def\vt{v_{m,n+1}}
\def\vb{v_{m,n-1}}
\def\vtr{v_{m+1,n+1}}
\def\vbl{v_{m-1,n-1}}
\def\vtl{v_{m-1,n+1}}
\def\vbr{v_{m+1,n-1}}

\def\U{U_{m,n}}
\def\Ul{U_{m-1,n}}
\def\Ur{U_{m+1,n}}
\def\Ut{U_{m,n+1}}
\def\Ub{U_{m,n-1}}
\def\Utr{U_{m+1,n+1}}
\def\Ubl{U_{m-1,n-1}}
\def\Utl{U_{m-1,n+1}}
\def\Ubr{U_{m+1,n-1}}
\def\V{V_{m,n}}
\def\Vl{V_{m-1,n}}
\def\Vr{V_{m+1,n}}
\def\Vt{V_{m,n+1}}
\def\Vb{V_{m,n-1}}
\def\Vtr{V_{m+1,n+1}}
\def\Vbl{V_{m-1,n-1}}
\def\Vtl{V_{m-1,n+1}}
\def\Vbr{V_{m+1,n-1}}
\def\tr{{\rm tr}\,}

\def\a{\alpha}
\def\b{\beta}
\def\g{\gamma}
\def\d{\delta}
\def\ep{\epsilon}
\def\e{\varepsilon}
\def\z{\zeta}
\def\t{\theta}
\def\k{\kappa}
\def\l{\lambda}
\def\s{\sigma}
\def\f{\varphi}
\def\w{\omega}
\def\v{{\hbox{v}}}
\def\u{{\hbox{u}}}
\def\x{{\hbox{x}}}

\newcommand{\ie}{{\it i.e.}}
\newcommand{\cmod}[1]{ \vert #1 \vert ^2 }
\newcommand{\cmodn}[2]{ \vert #1 \vert ^{#2} }
\newcommand{\nhat}{\mbox{\boldmath$\hat n$}}
\nopagebreak[3]
\bigskip

\title{ \bf Supersymmetric version of a Gaussian irrotational compressible fluid flow}
\vskip 1cm

\bigskip
\author{
A.~M. Grundland\thanks{email address: grundlan@crm.umontreal.ca}
\\
Centre de Recherches Math{\'e}matiques, Universit{\'e} de Montr{\'e}al,\\
C. P. 6128, Succ.\ Centre-ville, Montr{\'e}al, (QC) H3C 3J7,
Canada\\ Universit\'{e} du Qu\'{e}bec, Trois-Rivi\`{e}res, CP500 (QC) G9A 5H7, Canada \acc A. J. Hariton\thanks{email
address: hariton@lns.mit.edu}
\\
Center for Theoretical Physics, Massachusetts Institute of Technology, \\
77 Massachusetts Avenue, 26-570,
Cambridge, MA 02139, United States\\} \date{}

\maketitle
\begin{abstract}
The Lie point symmetries and corresponding invariant solutions are obtained for a Gaussian, irrotational, compressible fluid flow. A supersymmetric extension of this model is then formulated through the use of a superspace and superfield formalism. The Lie superalgebra of this extended model is determined and a classification of its subalgebras is performed. The method of symmetry reduction is systematically applied in order to derive special classes of invariant solutions of the supersymmetric model. Several new types of algebraic, hyperbolic, multi-solitonic and doubly periodic solutions are obtained in explicit form.\\
\end{abstract}

Short Title: Supersymmetric version of a Gaussian fluid flow


\vspace{3mm}

\begin{center}MIT-CTP-3862\end{center}

\newpage

\section{Introduction}

The system of partial differential equations describing a steady, irrotational and compressible fluid flow in a plane \cite{Loewner} is given by
\begin{equation}
\begin{split}
u_y-v_x&=0,\\
(\rho u)_x+(\rho v)_y&=0,
\end{split}
\label{qwerty3}
\end{equation}
where $(u,v)$ are the Cartesian components of the fluid velocity, and the density $\rho$ is defined as a function of $u$ and $v$. The fact that the fluid is irrotational (represented by the first of the two equations (\ref{qwerty3})) allows us to express the velocity field in terms of a potential function $\varphi$ such that $u=\varphi_x$, $v=\varphi_y$. Various density functions $\rho$ are admitted in the literature on the subject (see e.g. \cite{Lamb,Chandrasekhar,Jackiw}), and the most physically interesting instances involve bump type functions. One of the most well-investigated examples of the latter is 
\begin{equation}
\rho=(1+u^2+v^2)^{-1/2}.
\label{bidensity}
\end{equation}
When this density function $\rho$ is expressed in terms of the velocity potential $\varphi$ and substitued into the equations (\ref{qwerty3}), we obtain the minimal surfaces equation in $(2+1)$-dimensional Minkowski space \cite{Spivak}
\begin{equation}
\left(\varphi_x\over (1+(\varphi_x)^2+(\varphi_y)^2)^{1/2}\right)_x+\left(\varphi_y\over (1+(\varphi_x)^2+(\varphi_y)^2)^{1/2}\right)_y=0.
\label{qwerty2}
\end{equation}
In addition, by using the Wick rotation $y=it$, one may transform equation (\ref{qwerty2}) to the scalar Born-Infeld equation 
\begin{equation}
\left(1+(\varphi_x)^2\right)\varphi_{tt}-2\varphi_x\varphi_t\varphi_{xt}-\left(1-(\varphi_t)^2\right)\varphi_{xx}=0,
\label{qwerty1}
\end{equation}
which represents the simplest example of a nonlinear modification of Maxwell's electrodynamics in $(1+1)$ dimensions \cite{Arik}.

It is interesting to note that the Born-Infeld equation (\ref{qwerty1}) is compatible with the following hydrodynamical-type system expressed in terms of the Riemann invariants \cite{Jackiw}
\begin{equation}
R^+_t-R^-R^+_x=0,\qquad
R^-_t-R^+R^-_x=0.
\label{qwerty4}
\end{equation}
This can be shown by considering the transformation
\begin{equation}
R^{\pm}=\pm{(1+(\varphi_x)^2-(\varphi_t)^2)^{1/2}\over 1+(\varphi_x)^2}+{\varphi_x\varphi_t\over 1+(\varphi_x)^2}.
\label{qwerty5}
\end{equation}
Equation (\ref{qwerty1}) is also linked with the hyperbolic Monge-Amp\`{e}re equation \cite{Mokhov}
\begin{equation}
u_{xx}u_{tt}-(u_{xt})^2+1=0,
\label{qwerty6}
\end{equation}
via the Bianchi transformations
\begin{equation}
u_{tt}={(\varphi_t)^2-1\over (1-(\varphi_t)^2+(\varphi_x)^2)^{1/2}},\quad u_{xt}={\varphi_x\varphi_t\over (1-(\varphi_t)^2+(\varphi_x)^2)^{1/2}},\quad u_{tt}={(\varphi_x)^2+1\over (1-(\varphi_t)^2+(\varphi_x)^2)^{1/2}}.
\label{qwerty7}
\end{equation}
It should be noted that the system (\ref{qwerty4}) can be derived from the Monge-Amp\`{e}re equation (\ref{qwerty6}) if we substitute
\begin{equation}
R^{\pm}={1\over u_{xx}}(u_{xt}\pm 1).
\label{qwerty8}
\end{equation}
We may also proceed in the opposite direction by expressing $u_{xx}$ and $u_{xt}$ in terms of $R^+$ and $R^-$, and using the relations
\begin{equation}
u_{tt}={2R^+R^-\over R^+-R^-},\quad u_{xt}={R^++R^-\over R^+-R^-},\quad u_{xx}={2\over R^+-R^-}.
\label{qwerty9}
\end{equation}
It should be noted that the Monge-Amp\`{e}re equation (\ref{qwerty6}) can be linearized through a half-Legendre transformation \cite{MoserVeselov}
\begin{equation}
\tilde{u}(z,y)=u(s,y)-su_s(s,y),
\label{halflegendre}
\end{equation}
where we define
\begin{equation}
z=u_s,
\label{halflegendre2}
\end{equation}
provided that $u_{ss}\neq 0$. This transformation leads us directly to the linear wave equation
\begin{equation}
\tilde{u}_{yy}-\tilde{u}_{zz}=0.
\label{halflegendre3}
\end{equation}
Finally, the system for the Chaplygin gas
\begin{equation}
U_t=\frac{1}{2}\left(U^2-V^{-2}\right)_x,\qquad
V_t=\left(UV\right)_x,
\label{qwerty10}
\end{equation}
expressed as a conservation law, can be linked to the system (\ref{qwerty4}) through the relations
\begin{equation}
R^{\pm}=U\pm{1\over V},
\label{qwerty11}
\end{equation}
and to the Monge-Amp\`{e}re equation (\ref{qwerty6}) through the compatibility
\begin{equation}
U={u_{xt}\over u_{xx}},\quad V=u_{xx},
\label{qwerty12}
\end{equation}
along with the relation
\begin{equation}
u_{tt}=U^2V-V^{-1}.
\label{qwerty13}
\end{equation}
In fact, both systems (\ref{qwerty1}) and (\ref{qwerty10}) can be derived, through different parametrizations, from the Nambu-Goto action for a string evolving in a $(2+1)$-dimensional target space-time \cite{Jackiw}. The Chaplygin gas (\ref{qwerty10}) in $(1+1)$ dimensions is derived from the light cone gauge, while the Born-Infeld equation (\ref{qwerty1}) is obtained from a Cartesian parametrization. The symmetry properties, subalgebra classifications and invariant solutions of these models have been studied extensively \cite{Jackiw,GIS,PIS}.

In the past few years, there has also been a considerable amount of interest in the study of theories involving odd (fermionic) Grassmann variables and superalgebras. Such systems are interesting because ordinary matter generally consists of fermions wheras bosons are only concerned with the interactions. 
One relatively recent line of inquiry in this area of research has involved the construction of supersymmetric extensions of existing classical and quantum systems. This approach, first used in the context of particle physics, was successfully employed to supersymmetrize theories involving classical and quantum fields \cite{Jackiw,Manin,Kac}. Application of the techniques to fluid dynamics began with the study of simple Euler-type systems \cite{Fatyga,Roelofs,Hij}, followed by extensive work on a supersymmetric version of the Korteweg-de Vries equation \cite{Mathieu,Labelle,HusAyaWin}. More recently, the Chaplygin gas was supersymmetrized in both $(1+1)$ and $(2+1)$ dimensions by R. Jackiw, Y. Bergner and A. P. Polychronakos \cite{Jackiw, Bergner, Polychronakos} through the addition of fermionic-valued fields $\psi$ to the classical theory in the bosonic field $\theta$. In those models, the velocity was no longer irrotational but was expressed in terms of both the bosonic and fermionic fields
\begin{equation}
\mathbf{v}=\mathbf{\nabla}\theta-\frac{1}{2}\psi\mathbf{\nabla}\psi.
\label{polybergeq}
\end{equation}
Their approach involving a Lagrangian formulation of the fluid dynamics equations has been applied, among other areas, to an $N=1$ supersymmetric extension of polytropic gas dynamics \cite{Das}, a covariant and supersymmetric theory of relativistic hydrodynamics in four-dimensional Minkowski space \cite{Nyawelo1,Nyawelo2}, and a Kaluza-Klein model of a relativistic fluid \cite{Hassaine}.


It must be said that the physical interpretation of the systems resulting from the supersymmetric extensions remains an intriguing and still open question. However, a number of remarkable, physically meaningful solutions obtained in both particle physics \cite{Manin,Dunne} and fluid dynamics \cite{Nair} attest to the potential of this methodology and motivate further attempts at its application. In this paper, we apply it to a Gaussian fluid flow which, to our knowledge, has not been supersymmetrized before.



The objective of this paper is to investigate the equations of a steady, irrotational and compressible fluid flow (\ref{qwerty3}) involving the Gaussian density function
\begin{equation}
\rho=e^{-u^2-v^2}.
\label{nonsense1}
\end{equation}
This type of Gaussian fluid 
plays an essential role in many areas of physics, including fluid dynamics \cite{Chandrasekhar,Sommerfeld}, plasma physics \cite{Chen} and astrophysics \cite{Chandrasekhar}. 
When the density function (\ref{nonsense1}) is substitued into the equations (\ref{qwerty3}) and the velocity components expressed in terms of the potential $\varphi$, we obtain the equation
\begin{equation}
\left(1-(\varphi_x)^2\right)\varphi_{xx}-2\varphi_x\varphi_y\varphi_{xy}+\left(1-(\varphi_y)^2\right)\varphi_{yy}=0.
\label{qwertynot}
\end{equation}
Through the use of a method similar to that used for the Born-Infeld equation (\ref{qwerty1}), we will formulate a supersymmetric generalization of equation (\ref{qwertynot}). Due to its interesting symmetry properties and solutions, we will also consider a slightly modified version of equation (\ref{qwertynot}) found in statics of liquids \cite{Plateau}.
\begin{equation}
\left(1+(\varphi_x)^2\right)\varphi_{xx}-2\varphi_x\varphi_y\varphi_{xy}+\left(1+(\varphi_y)^2\right)\varphi_{yy}=0.
\label{qwertynot2}
\end{equation}
For the purpose of considering both equations (\ref{qwertynot}) and (\ref{qwertynot2}), let us introduce the parameter $\varepsilon=\pm 1$. The values $\varepsilon=1$ and $\varepsilon=-1$ will correspond to equations (\ref{qwertynot}) and (\ref{qwertynot2}) respectively. Thus, the two equations may be represented in the form
\begin{equation}
\left(1-\varepsilon(\varphi_x)^2\right)\varphi_{xx}-2\varphi_x\varphi_y\varphi_{xy}+\left(1-\varepsilon(\varphi_y)^2\right)\varphi_{yy}=0.
\label{qwertynot3}
\end{equation}
The goal is to construct supersymmetric extensions of both the Gaussian fluid flow equation (\ref{qwertynot}) and its modified version (\ref{qwertynot2}), to study their symmetry properties and to obtain new classes of invariant solutions of the extended models.
The approach used in this paper is different from that used by Jackiw, Bergner and Polychronakos for the Chaplygin gas, and is based on the symmetry reduction method adapted to Grassmann-valued partial differential equations. It has been used in the past for the supersymmetric Korteweg-de Vries equation \cite{HusAyaWin} and, more recently, by one of the authors for the supersymmetric Born-Infeld scalar model \cite{Hariton}.

Our paper is organized as follows. In section 2, we identify the symmetries, subalgebra classifications and invariant solutions of the classical Gaussian fluid flow equation (\ref{qwertynot3}). In section 3, we describe the most general form of the supersymmetric extension in terms of a superspace formalism. In section 4, we examine the symmetry properties of the extended system and compare them to those of the supersymmetric scalar Born-Infeld equation constructed previously and describe the subalgebra classification structure. In section 5, we use the symmetry reduction method to obtain invariant solutions, including elementary solutions (algebraic with one and two simple poles, trigonometric and hyperbolic) and doubly periodic solutions which can be expressed in terms of elliptic functions. Finally, section 6 contains a summary of our results.

\section{Symmetry properties of the classical Gaussian fluid flow}


Before we construct our supersymmetric extension, we first examine the Lie point symmetries and invariant solutions of the classical irrotational Gaussian fluid flow (\ref{qwertynot3}). 

\subsection{The case where $\varepsilon=1$}

We consider first of all the case where $\varepsilon=1$. The Lie symmetry algebra ${\mathcal G}_{(1)}$ of equation (\ref{qwertynot}) is spanned by the following five generators
\begin{equation}
\begin{split}
& S_{(1)}=x\partial_x+y\partial_y+\varphi\partial_{\varphi},\qquad M=-y\partial_x+x\partial_y,\\ & T_1=\partial_x,\qquad T_2=\partial_y,\qquad T_3=\partial_{\varphi},
\end{split}
\label{whoknowsnow1}
\end{equation}
which represent a dilation, a rotation and three translations in each of the three variables. The commutation relations of the algebra ${\mathcal G}_{(1)}$ spanned by the generators (\ref{whoknowsnow1}) are summarized in table 1.

\begin{table}[htbp]
\caption{Commutation table for the Lie symmetry algebra ${\mathcal G}_{(1)}$ generated by the vector fields (\ref{whoknowsnow1})}
\begin{center}
\begin{tabular}{|c||c|c|c|c|c|}\hline
 & $\mathbf{S_{(1)}}$ & $\mathbf{M}$ & $\mathbf{T_1}$ & $\mathbf{T_2}$ & $\mathbf{T_3}$ \\\hline\hline
$\mathbf{S_{(1)}}$ & $0$ & $0$ & $-T_1$ & $-T_2$ & $-T_3$ \\\hline
$\mathbf{M}$ & $0$ & $0$ & $-T_2$ & $T_1$ & $0$ \\\hline
$\mathbf{T_1}$ & $T_1$ & $T_2$ & $0$ & $0$ & $0$ \\\hline
$\mathbf{T_2}$ & $T_2$ & $-T_1$ & $0$ & $0$ & $0$ \\\hline
$\mathbf{T_3}$ & $T_3$ & $0$ & $0$ & $0$ & $0$ \\\hline
\end{tabular}
\end{center}
\end{table}

The Lie algebra ${\mathcal G}_{(1)}$ can be decomposed as the semi-direct sum
\begin{equation}
{\mathcal G}_{(1)}=\{S_{(1)},M\}\sdir\{T_1,T_2,T_3\}\mbox{.}
\label{decomp101}
\end{equation}
We now classify the one-dimensional subalgebras of the Lie algebra ${\mathcal G}_{(1)}$ using the classification methods as described in \cite{Winternitz}. That is, we construct a list of representatives of conjugacy classes of subalgebras in such a way that each one-dimensional subalgebra is conjugate to one and only one element of the list. We focus on one-dimensional subalgebras with orbits of codimension $1$ because we are only interested in invariant (as opposed to partially invariant, conditionally invariant or generic) solutions, so that one-dimensional reductions will lead to ordinary differential equations. For each of the elements of the following list, we first identify the representative subalgebra of the conjugacy class. Next, we describe the invariants of the subalgebra and (except in the case of $L_1$) the change of variable. Finally, using the symmetry reduction method as described in \cite{Olver,Clarkson}, we substitute the change of variable into the original equation (\ref{qwertynot}) in order to obtain a reduced ordinary differential equation in the symmetry variable $\xi$.
\begin{eqnarray}
L_1&=&\{T_3\},\qquad\mbox{Invariants: }x,\quad y;\nonumber
\end{eqnarray}
\begin{eqnarray}
L_2&=&\{T_1\},\qquad\mbox{Invariants: }\xi=y,\quad\varphi,\qquad\mbox{Change of variable: }\varphi=\varphi(y),\nonumber\\
& & \left(1-(\varphi_y)^2\right)\varphi_{yy}=0;
\end{eqnarray}
\begin{eqnarray}
L_3&=&\{T_3+mT_1\}_{m\neq 0},\qquad\mbox{Invariants: }\xi=y,\quad\varphi-{1\over m}x,\nonumber\\
& & \mbox{Change of variable: }\varphi=F(y)+{1\over m}x,\nonumber\\
& & \left(1-(F_y)^2\right)F_{yy}=0;
\end{eqnarray}
\begin{eqnarray}
L_4&=&\{S_{(1)}\},\qquad\mbox{Invariants: }\xi={x\over y},\quad{\varphi\over y},\qquad\mbox{Change of variable: }\varphi=yF(\xi),\nonumber\\
& & \left[(1+\xi^2)-(1+\xi^2)^2(F_{\xi})^2+2\xi(1+\xi^2)FF_{\xi}-\xi^2F^2\right]F_{\xi\xi}=0;\label{redeqno3}
\end{eqnarray}
\begin{eqnarray}
L_5&=&\{M\},\qquad\mbox{Invariants: }\xi=x^2+y^2,\quad\varphi,\qquad\mbox{Change of variable: }\varphi=\varphi(\xi),\nonumber\\
& & \varphi_{\xi}+\xi\varphi_{\xi\xi}-2\xi(\varphi_{\xi})^3-4\xi^2(\varphi_{\xi})^2\varphi_{\xi\xi}=0;\label{redeqno4}
\end{eqnarray}
\begin{eqnarray}
L_6&=&\{S_{(1)}+aM\}_{a\neq 0},\quad\mbox{Invariants: }\xi=\arctan{\left(y\over x\right)}-a\ln{\left(\sqrt{x^2+y^2}\right)},\quad{\varphi\over \sqrt{x^2+y^2}},\nonumber\\
& & \mbox{Change of variable: }\varphi=\sqrt{x^2+y^2}F(\xi),\nonumber\\
& & -2a(1+a^2)FF_{\xi}F_{\xi\xi}+(1+a^2)^2(F_{\xi})^2F_{\xi\xi}+a^2F^2F_{\xi\xi}-(1+a^2)F_{\xi\xi}\label{redeqno5}\\
& & -a(1+a^2)(F_{\xi})^3+(1+2a^2)F(F_{\xi})^2-aF^2F_{\xi}+2aF_{\xi}-F=0;\nonumber
\end{eqnarray}
\begin{eqnarray}
L_7&=&\{M+\mu T_3\}_{\mu\pm 1},\quad\mbox{Invariants: }\xi=x^2+y^2,\quad\varphi-\mu\arcsin{\left({y\over \sqrt{x^2+y^2}}\right)},\nonumber\\
& & \mbox{Change of variable: }\varphi=F(\xi)+\mu\arcsin{\left({y\over \sqrt{x^2+y^2}}\right)},\nonumber\\
& & F_{\xi}+2\xi F_{\xi}-4\xi^2(F_{\xi})^3+2\xi^2F_{\xi\xi}-8\xi^3(F_{\xi})^2F_{\xi\xi}=0.\label{redeqno6}
\label{algebraclasses}
 \end{eqnarray}
In the case of the subalgebra $L_1$, we obtain no reduced equation and therefore no invariant solution. This corresponds to the fact that there is no function $\varphi(x,y)$ which is invariant under the transformation $\varphi\longrightarrow\varphi+K$, where $K$ is a constant. Solutions invariant under the subalgebras $L_2$ and $L_3$ are linear polynomials in $x$ and $y$.\\\\ 
For subalgebra $L_4$, we obtain the following three solutions for equation (\ref{redeqno3}) 
\begin{equation}
F(\xi)=C_1\xi+C_2,
\end{equation}
which is a simple linear function, and the two expressions
\begin{equation}
F(\xi)=\sqrt{1+\xi^2}\left(\pm\arctan{\xi}+C_1\right),
\label{kinksolution}
\end{equation}
which correspond to a kink-type fluid density (a condensation wave).
For the subalgebra $L_6$, the solution of equation (\ref{redeqno4}) is
\begin{equation}
\phi(\xi)=\pm{i\over 2}\int\left({{\mathcal L}\left({-4C_1^2\over\xi}\right)\over \xi}\right)^{1/2}d\xi+C_2,
\label{lambertsolution}
\end{equation}
where the Lambert function, $y={\mathcal L}(x)$, is the solution to the equation 
\begin{equation}
ye^y=x.
\label{Lamberteqfn}
\end{equation}
Although, for each $x$, there are infinitely many values which satisfy (\ref{Lamberteqfn}), it should be noted that exactly one of the branches is analytic at zero. The multi-valuedness of this solution exhibits ergodic behavior \cite{Luban}.\\\\
 Finally, for subalgebras $L_6$ and $L_7$, equations (\ref{redeqno5}) and (\ref{redeqno6}) do not possess the Painlev\'{e} property.

\subsection{The case where $\varepsilon=-1$}

Considering now the case where $\varepsilon=-1$, we see that
the Lie symmetry algebra ${\mathcal G}_{(-1)}$ of equation (\ref{qwertynot2}) is spanned by the following four generators
\begin{equation}
\begin{split}
& S_{(-1)}=x\partial_x+y\partial_y+\varphi\partial_{\varphi},\qquad t_1=\partial_x,\qquad t_2=\partial_y,\qquad t_3=\partial_{\varphi},
\end{split}
\label{whoknows1}
\end{equation}
which differs from the case $\varepsilon=1$ in the sense that there is no rotation present. It should be noted that this algebra is solvable. A classification of the one-dimensional subalgebras of ${\mathcal G}_{(-1)}$ leads us to conclude that each subalgebra of the form $\{at_1+bt_2+ct_3\}$ (where $a$, $b$ and $c$ are real constants) is in its own separate conjugacy class, whereas every other subalgebra is conjugate to $\{S_{(-1)}\}$. Using the classical symmetry reduction method, one can obtain the following invariant solutions of (\ref{qwertynot2}). 
A solution invariant under the action of the dilation $S_{(-1)}$ must be of the form $\varphi=yF(\xi)$, where $\xi=x/y$ is the symmetry variable. In this case, equation (\ref{qwertynot2}) reduces to the ordinary differential equation
\begin{equation}
\begin{split}
&\left((1+\xi^2)+(1-\xi^2)^2(F_{\xi})^2+2\xi(1-\xi^2)FF_{\xi}+\xi^2F^2\right)F_{\xi\xi}=0.
\end{split}
\label{whoknows2}
\end{equation}
Due to the factorization of (\ref{whoknows2}), one obtains directly a solution which is linear in $\xi$ and corresponds to the (linear) solution $u(x,y)=C_1x+C_2y$, where $C_1$ and $C_2$ are real constants. In addition, one obtains the following two solutions expressed in terms of elliptic functions
\begin{equation}
\begin{split}
F(\xi)=&C_1\sqrt{\xi-1}\sqrt{\xi+1}\\ & \pm{\sqrt{-1-\xi^2}\over 1+\xi^2}\left(\xi+\xi^3+\sqrt{1-\xi^2}\sqrt{1+\xi^2}({\mathcal F}(\xi,i)-{\mathcal E}(\xi,i))\right),
\end{split}
\label{eqebis3ZZ}
\end{equation}
where
\begin{equation}
{\mathcal F}(z,k)=\int\limits_0^z{1\over \sqrt{1-\alpha^2}\sqrt{1-k^2\alpha^2}}d\alpha
\label{eqebis4ZZ}
\end{equation}
is the incomplete elliptic integral of the first kind (the inverse of the Jacobi function $sn$) and
\begin{equation}
{\mathcal E}(z,k)=\int\limits_0^z{\sqrt{1-k^2\alpha^2}\over \sqrt{1-\alpha^2}}d\alpha
\label{eqebis5ZZ}
\end{equation}
is the incomplete elliptic integral of the second kind \cite{Abram}.
Solutions invariant under a subalgebra of the form $\{at_1+bt_2+ct_3\}$ all consist of linear polynomials in $x$ and $y$.

\section{Supersymmetric extension}

In analogy with the case of the supersymmetric Born-Infeld scalar equation, we construct a supersymmetric extension of equations (\ref{qwertynot3}) through a superspace Grassmannian formalism. More specifically, we supplement the set of bosonic (even) independent variables $\{(x,y)\}$ with a fermionic (odd) Grassmann parameter $\theta$ and replace the bosonic function $\varphi(x,y)$ with the superfield $\Phi(x,y,\theta)$ defined as
\begin{equation}
\Phi(x,y,\theta) = \psi(x,y) + \theta\varphi(x,y)\mbox{,}
\label{eqb2}
\end{equation}
where $\psi(x,y)$ is a fermionic field. 
We construct our extension in such a way that it is invariant under the supersymmetry transformation
\begin{equation}
 x\rightarrow x-\underline{\eta}\theta\mbox{,}\qquad \theta\rightarrow\theta+\underline{\eta}\mbox{,}
\label{eqb3}
\end{equation}
where $\underline{\eta}$ is a constant fermionic parameter. This transformation is generated by the infinitesimal supersymmetry operator
\begin{equation}
H=\partial_{\theta}-\theta\partial_x\mbox{.}
\label{eqb4}
\end{equation}
In order to make our superfield theory manifestly invariant under the action of $H$, we write the equation in terms of the covariant derivative operator
\begin{equation}
D=\partial_{\theta}+\theta\partial_x\mbox{,}
\label{eqb5}
\end{equation}
which possesses the property that it anticommutes with the operator $H$. That is,
\begin{equation}
\{H,D\}=HD+DH=0.
\label{agaagnabooga1}
\end{equation}

The most general form of a supersymmetric extension of (\ref{qwertynot3}) is given by the following equation in terms of the superfield $\Phi$ and various orders and combinations of the derivatives $D$ and $\partial_y$
\begin{equation}
\begin{split}
& D^4\Phi -\varepsilon a(D^2\Phi)(D^3\Phi)(D^5\Phi) -\varepsilon (1-a)(D^3\Phi)^2(D^4\Phi) - 2b(D^2\Phi)(D\Phi)_y(D^3\Phi)_y\\ & - 2c(D^3\Phi)\Phi_y(D^3\Phi)_y - 2(1-b-c)(D^3\Phi)(D\Phi)_y(D^2\Phi)_y + \Phi_{yy}\\ & -\varepsilon d\Phi_y(D\Phi)_y(D\Phi)_{yy} -\varepsilon (1-d)\left((D\Phi)_y\right)^2\Phi_{yy} = 0\mbox{,}
\end{split}
\label{eqb6}
\end{equation}
where $a$, $b$, $c$ and $d$ are arbitrary real parameters. Here, the subscript $y$ indicates differentiation with respect to $y$. This equation can be decomposed into the following two partial differential equations for the fields $\varphi(x,y)$ and $\psi(x,y)$ corresponding respectively to the coefficient of the fermionic variable $\theta$ and that of the remaining terms
\begin{equation}
\begin{split}
& \varphi_{xx}-\varepsilon(\varphi_x)^2\varphi_{xx}+\varepsilon a\varphi_{xx}\psi_x\psi_{xx}+\varepsilon a\varphi_x\psi_x\psi_{xxx}-2\varphi_x\varphi_y\varphi_{xy}+2b\varphi_{xy}\psi_x\psi_{xy}+2b\varphi_y\psi_x\psi_{xxy}\\ & +2c\varphi_{xy}\psi_y\psi_{xx}+2c\varphi_x\psi_y\psi_{xxy}-2\varphi_y\psi_{xx}\psi_{xy}+2b\varphi_y\psi_{xx}\psi_{xy}+2c\varphi_y\psi_{xx}\psi_{xy}+\varphi_{yy}\\ & -\varepsilon(\varphi_y)^2\varphi_{yy}+\varepsilon d\varphi_y\psi_y\psi_{xyy}+\varepsilon d\varphi_{yy}\psi_y\psi_{xy}-2\varepsilon\varphi_y\psi_{xy}\psi_{yy}+2\varepsilon d\varphi_y\psi_{xy}\psi_{yy}=0\mbox{,}
\end{split}
\label{eqb7}
\end{equation}
and
\begin{equation}
\begin{split}
& \psi_{xx}-\varepsilon a\varphi_x\varphi_{xx}\psi_x-\varepsilon(\varphi_x)^2\psi_{xx}+\varepsilon a(\varphi_x)^2\psi_{xx}-2b\varphi_y\varphi_{xy}\psi_x-2c\varphi_x\varphi_{xy}\psi_y-2\varphi_x\varphi_y\psi_{xy}\\ & +2b\varphi_x\varphi_y\psi_{xy}+2c\varphi_x\varphi_y\psi_{xy}+\psi_{yy}-\varepsilon d\varphi_y\varphi_{yy}\psi_y-\varepsilon(\varphi_y)^2\psi_{yy}+\varepsilon d(\varphi_y)^2\psi_{yy}=0\mbox{.}
\end{split}
\label{eqb8}
\end{equation}
In the limit where $\psi\rightarrow 0$, equation (\ref{eqb7}) reduces to the standard bosonic equation (\ref{qwertynot3}). Equation (\ref{eqb8}) is a new equation involving the fermionic field $\psi$.

\section{Lie symmetry superalgebra}

For the system of partial differential equations defined by the system (\ref{eqb7}) and (\ref{eqb8}), the following symmetries are present for all values of the parameters $a$, $b$, $c$ and $d$. We have the following dilation in independent and dependent variables
\begin{equation}
S=x\partial_x+y\partial_y+\varphi\partial_{\varphi}+\frac{3}{2}\psi\partial_{\psi}\mbox{,}
\label{eqc1}
\end{equation}
along with the following four translations
\begin{equation}
P_1=\partial_x\mbox{,}\qquad P_2=\partial_y\mbox{,}\qquad Z=\partial_{\varphi}\mbox{,}\qquad Y=\partial_{\psi}\mbox{.}
\label{eqc2}
\end{equation}
In addition, in the case where $a=0$ and $b=0$, we have the following additional fermionic symmetry
\begin{equation}
Q_1=x\partial_{\psi}\mbox{.}
\label{eqc3}
\end{equation}
Similarily, in the case where $c=0$ and $d=0$, we have the additional generator
\begin{equation}
Q_2=y\partial_{\psi}\mbox{.}
\label{eqc4}
\end{equation}
Finally, it should be noted that the generators $H$ and $D$ given in equations (\ref{eqb4}) and (\ref{eqb5}), which involve the independent variables $x$ and $\theta$ in the superspace, correspond to the following generalized symmetries of the coordinate space $X\times U=\{(x,y,\varphi,\psi)\}$
\begin{equation}
H=-\psi_x\partial_{\varphi}+\varphi\partial_{\psi},
\label{qqq}
\end{equation}
and
\begin{equation}
D=\psi_x\partial_{\varphi}+\varphi\partial_{\psi}.
\label{ddd}
\end{equation}

It is interesting to note that the symmetry generators found for our supersymmetric Gaussian fluid flow equations (\ref{eqb7}) and (\ref{eqb8}) are analogous to those found for the supersymmetric scalar Born-Infeld extension \cite{Hariton}. The only instance for which all seven symmetries (\ref{eqc1}), (\ref{eqc2}), (\ref{eqc3}) and (\ref{eqc4}) are present is the case where $a=0$, $b=0$, $c=0$ and $d=0$. From here onwards, we shall focus our attentions on this specific case. The equations (\ref{eqb6}), (\ref{eqb7}) and (\ref{eqb8}) become
\begin{equation}
\begin{split}
& D^4\Phi -\varepsilon (D^3\Phi)^2(D^4\Phi) - 2(D^3\Phi)(D\Phi)_y(D^2\Phi)_y + \Phi_{yy} -\varepsilon \left((D\Phi)_y\right)^2\Phi_{yy} = 0\mbox{,}
\end{split}
\label{eqc5}
\end{equation}
\begin{equation}
\begin{split}
& \varphi_{xx}-\varepsilon(\varphi_x)^2\varphi_{xx}-2\varphi_x\varphi_y\varphi_{xy}-2\varphi_y\psi_{xx}\psi_{xy}+\varphi_{yy} -\varepsilon(\varphi_y)^2\varphi_{yy}-2\varepsilon\varphi_y\psi_{xy}\psi_{yy}=0\mbox{,}
\end{split}
\label{eqc6}
\end{equation}
and
\begin{equation}
\begin{split}
& \psi_{xx}-\varepsilon(\varphi_x)^2\psi_{xx}-2\varphi_x\varphi_y\psi_{xy}+\psi_{yy}-\varepsilon(\varphi_y)^2\psi_{yy}=0\mbox{,}
\end{split}
\label{eqc7}
\end{equation}
respectively. The commutation and anticommutation relations of the superalgebra $L$ formed by the generators (\ref{eqc1}), (\ref{eqc2}), (\ref{eqc3}) and (\ref{eqc4}) are summarized in table 2.

\begin{table}[htbp]
\caption{Commutation table for the Lie symmetry superalgebra in the case where $a,b,c,d=0$}
\begin{center}
\begin{tabular}{|c||c|c|c|c|c|c|c|}\hline
 & $\mathbf{S}$ & $\mathbf{P_1}$ & $\mathbf{P_2}$ & $\mathbf{Z}$ & $\mathbf{Y}$ & $\mathbf{Q_1}$ & $\mathbf{Q_2}$ \\\hline\hline
$\mathbf{S}$ & $0$ & $-P_1$ & $-P_2$ & $-Z$ & $-\frac{3}{2}Y$ & $-\frac{1}{2}Q_1$ & $-\frac{1}{2}Q_2$ \\\hline
$\mathbf{P_1}$ & $P_1$ & $0$ & $0$ & $0$ & $0$ & $Y$ & $0$ \\\hline
$\mathbf{P_2}$ & $P_2$ & $0$ & $0$ & $0$ & $0$ & $0$ & $Y$ \\\hline
$\mathbf{Z}$ & $Z$ & $0$ & $0$ & $0$ & $0$ & $0$ & $0$ \\\hline
$\mathbf{Y}$ & $\frac{3}{2}Y$ & $0$ & $0$ & $0$ & $0$ & $0$ & $0$ \\\hline
$\mathbf{Q_1}$ & $\frac{1}{2}Q_1$ & $-Y$ & $0$ & $0$ & $0$ & $0$ & $0$ \\\hline
$\mathbf{Q_2}$ & $\frac{1}{2}Q_2$ & $0$ & $-Y$ & $0$ & $0$ & $0$ & $0$ \\\hline
\end{tabular}
\end{center}
\end{table}

\subsection{One-dimensional subalgebras}

In this section, we describe the classification of the one-dimensional subalgebras of the Lie superalgebra $L$ obtained in section 3. In other words, we proceed to construct a list of representatives of conjugacy classes of subalgebras in such a way that each one-dimensional subalgebra is conjugate to one and only one element of the list. This classification has already been performed for the Born-Infeld superalgebra \cite{Hariton}, and so we describe the results. We focus on one-dimensional subalgebras with orbits of codimension $1$ because we are only interested in invariant (as opposed to partially invariant, conditionally invariant or generic) solutions, so that one-dimensional reductions will lead to ordinary differential equations. The analysis led to the following results.

The one-dimensional splitting subalgebras are
\begin{equation}
\begin{split}
& L_1=\{S\}\mbox{,}\qquad L_2=\{P_1\}\mbox{,}\qquad L_3=\{P_2\}\mbox{,}\qquad L_{4,m}=\{P_1+mP_2\}_{m\neq 0}\mbox{,}\\ & L_5=\{Z\}\mbox{,}\qquad L_{6,m}=\{Z+mP_1\}_{m\neq 0}\mbox{,}\qquad L_{7,m}=\{Z+mP_2\}_{m\neq 0}\mbox{,}\\ & L_{8,m,n}=\{Z+mP_1+nP_2\}_{m,n\neq 0}\mbox{,}\qquad L_9=\{Y\}\mbox{,}\qquad L_{10}=\{Q_1\}\mbox{,}\qquad L_{11}=\{Q_2\}\mbox{,}\\ & L_{12,k}=\{Q_1+kQ_2\}_{k\neq 0}\mbox{,}
\end{split}
\label{eqd1}
\end{equation}
and the non-splitting subalgebras are
\begin{equation}
\begin{split}
& {\mathcal L}_{2,\underline{\eta_1},\underline{\eta_2}}=\{P_1+\underline{\eta_1}Q_1+\underline{\eta_2}Q_2\}\mbox{,}\qquad {\mathcal L}_{3,\underline{\eta_1},\underline{\eta_2}}=\{P_2+\underline{\eta_1}Q_1+\underline{\eta_2}Q_2\}\mbox{,}\\ & {\mathcal L}_{4,m,\underline{\eta_1},\underline{\eta_2}}=\{P_1+mP_2+\underline{\eta_1}Q_1+\underline{\eta_2}Q_2\}_{m\neq 0}\mbox{,}\qquad {\mathcal L}_{5,\underline{\eta_1},\underline{\eta_2}}=\{Z+\underline{\eta_1}Q_1+\underline{\eta_2}Q_2\}\mbox{,}\\ & {\mathcal L}_{6,m,\underline{\eta_1},\underline{\eta_2}}=\{Z+mP_1+\underline{\eta_1}Q_1+\underline{\eta_2}Q_2\}_{m\neq 0}\mbox{,}\\ & {\mathcal L}_{7,m,\underline{\eta_1},\underline{\eta_2}}=\{Z+mP_2+\underline{\eta_1}Q_1+\underline{\eta_2}Q_2\}_{m\neq 0}\mbox{,}\\ & {\mathcal L}_{8,m,n,\underline{\eta_1},\underline{\eta_2}}=\{Z+mP_1+nP_2+\underline{\eta_1}Q_1+\underline{\eta_2}Q_2\}_{m,n\neq 0}\mbox{,}
\end{split}
\label{eqd2}
\end{equation}
where $m$, $n$ and $k$ are constant bosonic parameters, and $\underline{\eta_1}$ and $\underline{\eta_2}$ are fermionic constants. Throughout this paper, we use the convention that underlined constants are fermionic. It is important to note that each non-splitting subalgebra ${\mathcal L}_{i,\underline{\eta_1},\underline{\eta_2}}$ is an element of the same conjugacy class as the subalgebra ${\mathcal L}_{i,K\underline{\eta_1},K\underline{\eta_2}}$, where $K$ is a positive constant.

\section{Symmetry group reductions and solutions}

In this section, we make use of the classical symmetry reduction method to determine the invariants and reduced equations corresponding to each of the one-dimensional subalgebras in equations (\ref{eqd1}) and (\ref{eqd2}). In each case, we obtain at least one invariant expressed solely in terms of the independent variables $x$ and $y$, which we call the symmetry variable $\xi$. In most other cases, the remaining two invariants involve the dependent functions $\varphi$ and $\psi$, and they are identified as the functions $F(\xi)$ and $\Lambda(\xi)$ respectively. Expressing the fields $\varphi$ and $\psi$ in terms of the invariants, we substitute the derivatives into the supersymmetric system (\ref{eqc6}) and (\ref{eqc7}) in order to obtain the reduced a system of ordinary differential equations. The results are presented in tables 3 and 4. Where the Painlev\'{e} property is satisfied for these ODEs, we give an exact, analytic solution of the system of reduced equations, along with the corresponding solution of the supersymmetric system (\ref{eqd1}) and (\ref{eqd2}). It should be noted that this reduction procedure cannot be employed for subalgebras whose symmetry generators do not involve derivatives with respect to independent variables, nor for the generalized symmetries (\ref{qqq}) and (\ref{ddd}). Indeed, such transformations involve shifts only in the dependent variables, which cannot under any circumstances leave a solution invariant.

\begin{table}[htbp]
\caption{Invariants and change for variable for each splitting and non-splitting subalgebra}
\begin{center}
\begin{tabular}{|c|c|c|}\hline
Subalgebra & Invariants & relations and change of variable\\\hline\hline
$L_1=\{D\}$ & $\xi={x\over y},{\varphi\over y},y^{-3/2}\psi$ & $\varphi=yF(\xi),\psi=y^{3/2}\Lambda(\xi)$\\\hline
$L_2=\{P_1\}$ & $y,\varphi,\psi$ & $\varphi=\varphi(y),\psi=\psi(y)$\\\hline
$L_3=\{P_2\}$ & $x,\varphi,\psi$ & $\varphi=\varphi(x),\psi=\psi(x)$\\\hline
$L_{4,m}=\{P_1+mP_2\}_{m\neq 0}$ & $\xi=y-mx,\varphi,\psi$ & $\varphi=\varphi(\xi),\psi=\psi(\xi)$\\\hline
$L_5=\{Z\}$ & $x,y,\psi$ & $N/A$\\\hline
$L_{6,m}=\{Z+mP_1\}_{m\neq 0}$ & $y,x-m\varphi,\psi$ & $\varphi={1\over m}\left(F(y)+x\right),\psi=\psi(y)$\\\hline
$L_{7,m}=\{Z+mP_2\}_{m\neq 0}$ & $x,y-m\varphi,\psi$ & $\varphi={1\over m}\left(F(x)+y\right),\psi=\psi(x)$\\\hline
$L_{8,m,n}=$ & $\xi=x-{m\over n}y,\varphi-{y\over n},\psi$ & $\varphi=F(\xi)+{1\over n}y,\psi=\psi(\xi)$\\
$\{Z+mP_1+nP_2\}_{m,n\neq 0}$ & & \\\hline
$L_9=\{Y\}$ & $x,y,\varphi$ & $N/A$\\\hline
$L_{10}=\{Q_1\}$ & $x,y,\varphi$ & $N/A$\\\hline
$L_{11}=\{Q_2\}$ & $x,y,\varphi$ & $N/A$\\\hline
$L_{12,k}=\{Q_1+kQ_2\}_{k\neq 0}$ & $x,y,\varphi$ & $N/A$\\\hline
${\mathcal L}_{2,\underline{\eta_1},\underline{\eta_2}}=\{P_1$ & $y,\varphi,\psi-{1\over 2}\underline{\eta_1}x^2-\underline{\eta_2}xy$ & $\varphi=\varphi(y),\psi=\Lambda(y)+{1\over 2}\underline{\eta_1}x^2+\underline{\eta_2}xy$\\
$+\underline{\eta_1}Q_1+\underline{\eta_2}Q_2\}$ & & \\\hline
${\mathcal L}_{3,\underline{\eta_1},\underline{\eta_2}}=\{P_2$ & $x,\varphi,\psi-\underline{\eta_1}xy-{1\over 2}\underline{\eta_2}y^2$ & $\varphi=\varphi(x),\psi=\Lambda(x)+\underline{\eta_1}xy+{1\over 2}\underline{\eta_2}y^2$\\
$+\underline{\eta_1}Q_1+\underline{\eta_2}Q_2\}$ & & \\\hline
${\mathcal L}_{4,m,\underline{\eta_1},\underline{\eta_2}}=\{P_1+mP_2$ & $\xi=y-mx,\varphi,$ & $\varphi=\varphi(\xi),$\\
$+\underline{\eta_1}Q_1+\underline{\eta_2}Q_2\}_{m\neq 0}$ & $\psi-{1\over 2}\underline{\eta_1}x^2+{1\over 2}\underline{\eta_2}mx^2-\underline{\eta_2}xy$ & $\psi=\Lambda(\xi)+{1\over 2}\underline{\eta_1}x^2-{1\over 2}\underline{\eta_2}mx^2+\underline{\eta_2}xy$ \\\hline
${\mathcal L}_{5,\underline{\eta_1},\underline{\eta_2}}=\{Z$ & $x,y,\psi-\left(\underline{\eta_1}x+\underline{\eta_2}y\right)\varphi$ & $N/A$\\
$+\underline{\eta_1}Q_1+\underline{\eta_2}Q_2\}$ & & \\\hline
${\mathcal L}_{6,m,\underline{\eta_1},\underline{\eta_2}}=\{Z+mP_1$ & $y,\varphi-{1\over m}x,\psi-{1\over 2m}\underline{\eta_1}x^2-{1\over m}\underline{\eta_2}xy$ & $\varphi=F(y)+{1\over m}x,$\\
$+\underline{\eta_1}Q_1+\underline{\eta_2}Q_2\}_{m\neq 0}$ & & $\psi=\Lambda(y)+{1\over 2m}\underline{\eta_1}x^2+{1\over m}\underline{\eta_2}xy$ \\\hline
${\mathcal L}_{7,m,\underline{\eta_1},\underline{\eta_2}}=\{Z+mP_2$ & $x,\varphi-{1\over m}y,\psi-{1\over m}\underline{\eta_1}xy-{1\over 2m}\underline{\eta_2}y^2$ & $\varphi=F(x)+{1\over m}y,$\\
$+\underline{\eta_1}Q_1+\underline{\eta_2}Q_2\}_{m\neq 0}$ & & $\psi=\Lambda(x)+{1\over m}\underline{\eta_1}xy+{1\over 2m}\underline{\eta_2}y^2$ \\\hline
${\mathcal L}_{8,m,n,\underline{\eta_1},\underline{\eta_2}}=$ & $\xi=x-{m\over n}y,\varphi-{1\over n}y,$ & $\varphi=F(\xi)+{1\over n}y,$\\
$\{Z+mP_1+nP_2$ & $\psi-{1\over n}\underline{\eta_1}xy+{m\over 2n^2}\underline{\eta_1}y^2-{1\over 2n}\underline{\eta_2}y^2$ & $\psi=\Lambda(\xi)+{1\over n}\underline{\eta_1}xy-{m\over 2n^2}\underline{\eta_1}y^2+{1\over 2n}\underline{\eta_2}y^2$ \\
$+\underline{\eta_1}Q_1+\underline{\eta_2}Q_2\}_{m,n\neq 0}$ & & \\\hline
\end{tabular}
\end{center}
\end{table}

\begin{table}[htbp]
\caption{Reduced Equations corresponding to each splitting and non-splitting subalgebra}
\begin{center}
\begin{tabular}{|c|c|}\hline
Subalgebra & Reduced equations\\\hline\hline
$L_1=\{D\}$ & $(1+\xi^2)F_{\xi\xi}-\varepsilon(1+\varepsilon\xi^2)^2(F_{\xi})^2F_{\xi\xi}+2\xi(1+\varepsilon\xi^2)FF_{\xi}F_{\xi\xi}$ \\
 & $-\varepsilon\xi^2F^2F_{\xi\xi}+\frac{3}{4}\varepsilon(F-\xi F_{\xi})\Lambda\Lambda_{\xi}-\frac{3}{2}\varepsilon\xi(F-\xi F_{\xi})\Lambda\Lambda_{\xi\xi}$ \\
 & $+(1+\varepsilon\xi^2)(F-\xi F_{\xi})\Lambda_{\xi}\Lambda_{\xi\xi}=0$, \\
 & $(1+\xi^2)\Lambda_{\xi\xi}-\varepsilon(1+\varepsilon\xi^2)^2(F_{\xi})^2\Lambda_{\xi\xi}+2\xi(1+\varepsilon\xi^2)FF_{\xi}\Lambda_{\xi\xi}$ \\
 & $-\varepsilon\xi^2F^2\Lambda_{\xi\xi}+\xi(1+\varepsilon\xi^2)(F_{\xi})^2\Lambda_{\xi}-\varepsilon(2\xi^2+\varepsilon)FF_{\xi}\Lambda_{\xi}$ \\
 & $+\varepsilon\xi F^2\Lambda_{\xi}-\xi\Lambda_{\xi}-\frac{3}{4}\varepsilon\xi^2(F_{\xi})^2\Lambda+\frac{3}{2}\varepsilon\xi FF_{\xi}\Lambda-\frac{3}{4}\varepsilon F^2\Lambda+\frac{3}{4}\Lambda=0$ \\\hline
$L_2=\{P_1\}$ & $\varphi_{yy}-\varepsilon(\varphi_y)^2\varphi_{yy}=0$, \\
 & $\psi_{yy}-\varepsilon(\varphi_y)^2\psi_{yy}=0$ \\\hline
$L_3=\{P_2\}$ & $\varphi_{xx}-\varepsilon(\varphi_x)^2\varphi_{xx}=0$, \\
 & $\psi_{xx}-\varepsilon(\varphi_x)^2\psi_{xx}=0$ \\\hline
$L_{4,m}=\{P_1+mP_2\}_{m\neq 0}$ & $(m^2+1)\varphi_{\xi\xi}-\varepsilon(m^2+\varepsilon)^2(\varphi_{\xi})^2\varphi_{\xi\xi}=0$, \\
 & $(m^2+1)\psi_{\xi\xi}-\varepsilon(m^2+\varepsilon)^2(\varphi_{\xi})^2\psi_{\xi\xi}=0$ \\\hline
$L_{6,m}=\{Z+mP_1\}_{m\neq 0}$ & $m^2F_{yy}-\varepsilon(F_y)^2F_{yy}=0$, \\
 & $m^2\psi_{yy}-\varepsilon(F_y)^2\psi_{yy}=0$ \\\hline
$L_{7,m}=\{Z+mP_2\}_{m\neq 0}$ & $m^2F_{xx}-\varepsilon(F_x)^2F_{xx}=0$, \\
 & $m^2\psi_{xx}-\varepsilon(F_x)^2\psi_{xx}=0$ \\\hline
$L_{8,m,n}=\{Z+mP_1+nP_2\}_{m,n\neq 0}$ & $\left(-\varepsilon(1+\varepsilon{m^2\over n^2})^2(F_{\xi})^2+(2{m\over n^2}+2\varepsilon{m^3\over n^4})F_{\xi}+(1+{m^2\over n^2}-\varepsilon{m^2\over n^4})\right)F_{\xi\xi}=0$, \\
 & $\left(-\varepsilon(1+\varepsilon{m^2\over n^2})^2(F_{\xi})^2+(2{m\over n^2}+2\varepsilon{m^3\over n^4})F_{\xi}+(1+{m^2\over n^2}-\varepsilon{m^2\over n^4})\right)\psi_{\xi\xi}=0$ \\\hline
${\mathcal L}_{2,\underline{\eta_1},\underline{\eta_2}}=\{P_1$ & $-2\varepsilon \varphi_y\underline{\eta_2}\Lambda_{yy}-2\varphi_y\underline{\eta_1}\underline{\eta_2}+\varphi_{yy}-\varepsilon(\varphi_y)^2\varphi_{yy}=0,$ \\
$+\underline{\eta_1}Q_1+\underline{\eta_2}Q_2\}$ & $\Lambda_{yy}-\varepsilon(\varphi_y)^2\Lambda_{yy}+\underline{\eta_1}=0$ \\\hline
${\mathcal L}_{3,\underline{\eta_1},\underline{\eta_2}}=\{P_2$ & $\left(1-\varepsilon(\varphi_x)^2\right)\varphi_{xx}=0,$ \\
$+\underline{\eta_1}Q_1+\underline{\eta_2}Q_2\}$ & $\Lambda_{xx}-\varepsilon(\varphi_x)^2\Lambda_{xx}+\underline{\eta_2}=0$ \\\hline
${\mathcal L}_{4,m,\underline{\eta_1},\underline{\eta_2}}=\{P_1+mP_2$ & $(1+m^2)\varphi_{\xi\xi}-\varepsilon(1+\varepsilon m^2)^2(\varphi_{\xi})^2\varphi_{\xi\xi}$ \\
$+\underline{\eta_1}Q_1+\underline{\eta_2}Q_2\}_{m\neq 0}$ & $+2\varphi_{\xi}\left((-\varepsilon\underline{\eta_2}+m\underline{\eta_1})\Lambda_{\xi\xi}+\underline{\eta_2}\underline{\eta_1}\right)=0,$ \\
 & $(1+m^2)\Lambda_{\xi\xi}-\varepsilon(1+\varepsilon m^2)^2(\varphi_{\xi})^2\Lambda_{\xi\xi}$ \\
 & $+m(2\underline{\eta_2}-\varepsilon m\underline{\eta_1}+\varepsilon m^2\underline{\eta_2})(\varphi_{\xi})^2+(\underline{\eta_1}-m\underline{\eta_2})=0$ \\\hline
${\mathcal L}_{6,m,\underline{\eta_1},\underline{\eta_2}}=\{Z+mP_1$ & $F_{yy}-\varepsilon(F_y)^2F_{yy}-{2\over m^2}\underline{\eta_1}\underline{\eta_2}F_y-\varepsilon{2\over m}\underline{\eta_2}F_y\Lambda_{yy}=0,$ \\
$+\underline{\eta_1}Q_1+\underline{\eta_2}Q_2\}_{m\neq 0}$ & $\Lambda_{yy}-\varepsilon(F_y)^2\Lambda_{yy}-{2\over m^2}\underline{\eta_2}F_y-\varepsilon{1\over m^3}(1-\varepsilon m^2)\underline{\eta_1}=0$ \\\hline
${\mathcal L}_{7,m,\underline{\eta_1},\underline{\eta_2}}=\{Z+mP_2$ & $F_{xx}-\varepsilon(F_x)^2F_{xx}+{2\over m^2}\underline{\eta_1}\Lambda_{xx}-\varepsilon{2\over m^3}\underline{\eta_1}\underline{\eta_2}=0,$ \\
$+\underline{\eta_1}Q_1+\underline{\eta_2}Q_2\}_{m\neq 0}$ & $\Lambda_{xx}-\varepsilon(F_x)^2\Lambda_{xx}-{2\over m^2}\underline{\eta_1}F_x-\varepsilon{1\over m^3}(1-\varepsilon m^2)\underline{\eta_2}=0$ \\\hline
${\mathcal L}_{8,m,n,\underline{\eta_1},\underline{\eta_2}}=\{Z+mP_1+nP_2$ & $\left(-\varepsilon(1+\varepsilon{m^2\over n^2})^2(F_{\xi})^2+(2{m\over n^2}+2\varepsilon{m^3\over n^4})F_{\xi}+(1+{m^2\over n^2}-\varepsilon{m^2\over n^4})\right)F_{\xi\xi}$ \\
$+\underline{\eta_1}Q_1+\underline{\eta_2}Q_2\}_{m,n\neq 0}$ & $+{2\over n^2}\left(\underline{\eta_1}-\varepsilon{m\over n}\underline{\eta_2}\right)(1-mF_{\xi})\Lambda_{\xi\xi}-\varepsilon{2\over n^3}(1-mF_{\xi})\underline{\eta_1}\underline{\eta_2}=0,$ \\
 & $\left(-\varepsilon(1+\varepsilon{m^2\over n^2})^2(F_{\xi})^2+(2{m\over n^2}+2\varepsilon{m^3\over n^4})F_{\xi}+(1+{m^2\over n^2}-\varepsilon{m^2\over n^4})\right)\Lambda_{\xi\xi}$ \\
 & $+\left(-\varepsilon{2m^2\over n^4}F_{\xi}-{2\over n^2}F_{\xi}+{2m\over n^2}(F_{\xi})^2+\varepsilon{m^3\over n^4}(F_{\xi})^2-{m\over n^2}+\varepsilon{m\over n^4}\right)\underline{\eta_1}$ \\
 & $+\left({1\over n}-\varepsilon{1\over n^3}+\varepsilon{2m\over n^3}F_{\xi}-\varepsilon{m^2\over n^3}(F_{\xi})^2\right)\underline{\eta_2}=0$ \\\hline
\end{tabular}
\end{center}
\end{table}

\subsection{Splitting subalgebras and their reductions}

We begin our analysis by considering the splitting subalgebras. In the case of subalgebra $L_1$, we obtain two equations involving the functions $F$ and $\Lambda$, coupled in an involved way. In the specific case where the condition
\begin{equation}
\left(\frac{3}{4}\varepsilon\Lambda\Lambda_{\xi}-\frac{3}{2}\varepsilon\xi\Lambda\Lambda_{\xi\xi}+(1+\varepsilon\xi^2)\Lambda_{\xi}\Lambda_{\xi\xi}\right)(F-\xi F_{\xi})=0
\label{eqe1}
\end{equation}
holds, we can decouple the first equation which then becomes an ordinary differential equation in $F$
\begin{equation}
\left((1+\xi^2)-\varepsilon(1+\varepsilon\xi^2)^2(F_{\xi})^2+2\xi(1+\varepsilon\xi^2)FF_{\xi}-\varepsilon\xi^2F^2\right)F_{\xi\xi}=0\mbox{.}
\label{eqe2}
\end{equation}
In the case where $\varepsilon=1$, there are three functionally independent solutions of equation (\ref{eqe2}). The first two are expressed in terms of radicals and inverse trigonometric functions
\begin{equation}
F(\xi)=\sqrt{1+\xi^2}\left(\pm\arctan{\xi}+C_1\right)
\label{eqe3biss}
\end{equation}
Here, the function $\Lambda$ can be determined from the second reduced equation in combination with the condition (\ref{eqe1}), and is given by the expression
\begin{equation}
\begin{split}
\Lambda(\xi)=& \underline{E_1}\mbox{ }(1+\xi^2)^{3/4}\left(\pm\arctan{\xi}+C_1\right)^{3/4}\mbox{exp}\Bigg{(}-\left(\pm\frac{3}{4}C_1+\frac{3}{8}\arctan{\xi}\right)\arctan{\xi}\Bigg{)}.
\end{split}
\label{eqe6biss}
\end{equation}
Here, condition (\ref{eqe1}) is identically satisfied due to the fact that the constant $\underline{E_1}$ is fermionic. The solution (\ref{eqe3biss}) corresponds to a kink-type fluid density
\begin{equation}
\rho=e^{-\left(1+\left(\arctan{x\over y}+C_1\right)^2\right)},\qquad C_1\in \mathbb{R},
\label{a1}
\end{equation}
where the velocity components are given by
\begin{eqnarray}
u&={x\over \sqrt{x^2+y^2}}\left(\arctan{x\over y}+C_1\right)+{y\over \sqrt{x^2+y^2}}\\
v&={y\over \sqrt{x^2+y^2}}\left(\arctan{x\over y}+C_1\right)-{x\over \sqrt{x^2+y^2}}.
\label{a2}
\end{eqnarray}
This is a kink-type density solution
, which varies from the value $e^{-\left(1+\left(-{\Pi\over 2}+C_1\right)^2\right)}$ to the value $e^{-\left(1+\left({\Pi\over 2}+C_1\right)^2\right)}$, and depends only on the polar angle of the position on the plane. This solution is real, asymptotic and discontinuous on the $x$-axis, and represents a condensation wave.

The third solution of equation (\ref{eqe2}) is the linear function
\begin{equation}
F(\xi)=C_1\xi+C_2,
\label{eqe7biss}
\end{equation}
which gives us the following solution for $\Lambda$
\begin{equation}
\begin{split}
\Lambda(\xi)=& {\underline{E_1}\left((C_2^2-1)\xi^2-2C_1C_2\xi+(C_1^2-1)\right)^{3/2}\over \left((1-C_2^2)\xi+C_1C_2+\sqrt{C_1^2+C_2^2-1}\right)^{3/2}}.
\end{split}
\label{eqe999biss}
\end{equation}
In the case where $\varepsilon=-1$, there are again three independent solutions of equation (\ref{eqe2}). The first two are given in terms of elliptic functions
\begin{equation}
\begin{split}
F(\xi)=&C_1\sqrt{\xi-1}\sqrt{\xi+1}\pm\\ & {\sqrt{-1-\xi^2}\over 1+\xi^2}\left(\xi+\xi^3+\sqrt{1-\xi^2}\sqrt{1+\xi^2}({\mathcal F}(\xi,i)-{\mathcal E}(\xi,i))\right),
\end{split}
\label{eqe3}
\end{equation}
where $C_1\in\mathbb{R}$, and ${\mathcal F}(z,k)$ and ${\mathcal E}(z,k)$ are the incomplete elliptic integrals of the first and second kinds as defined in equations (\ref{eqebis4ZZ}) and (\ref{eqebis5ZZ}) respectively. Solution (\ref{eqe3}) is a complex-valued function, but for large positive values of $C_1$, the imaginary part provides a good approximation for a localized static bump solution.
The function $\Lambda$ 
is given by the expression
\begin{equation}
\begin{split}
\Lambda(\xi)=& \underline{E_1}\mbox{ }\mbox{exp}\Bigg{(}{3\over 4}\int\bigg{(}C_1\big{(}{\mathcal F}(\xi,i)-{\mathcal E}(\xi,i)\big{)}\big{(}2\xi^{10}+6\xi^8+4\xi^6-4\xi^4-6\xi^2-2\big{)}\\ & +\sqrt{-1-\xi^2}(1+\xi^2)^{7/2}\sqrt{\xi-1}\sqrt{\xi+1}\sqrt{(-\xi-1)(\xi-1)}\mbox{ }\cdot\\ & \big{(}1+C_1^2+\xi^2+({\mathcal F}(\xi,i)-{\mathcal E}(\xi,i))^2\big{)}\bigg{)}\bigg{/}\\ & \bigg{(}(\xi-1)(\xi+1)\Big{(}\big{(}{\mathcal E}(\xi,i)-{\mathcal F}(\xi,i)\big{)}\sqrt{-1-\xi^2}\sqrt{\xi-1}\sqrt{\xi+1}(1+\xi^2)^4\\ & +C_1(1+\xi^2)^{9/2}\sqrt{(-\xi-1)(\xi-1)}\\ & +x\sqrt{-1-\xi^2}(1+\xi^2)^{7/2}\sqrt{\xi-1}\sqrt{\xi+1}\sqrt{(-\xi-1)(\xi-1)}\Big{)}\bigg{)} d\xi\Bigg{)}.
\end{split}
\label{eqe6}
\end{equation}
The third solution of equation (\ref{eqe2}) is the linear function
\begin{equation}
F(\xi)=C_1\xi+C_2,
\label{eqe7}
\end{equation}
which gives us the following equation for $\Lambda$
\begin{equation}
(1+C_1^2+2C_1C_2\xi+\xi^2+C_2^2\xi^2)\Lambda_{\xi\xi}+(-C_2^2\xi-\xi-C_1C_2)\Lambda_{\xi}+\frac{3}{4}(1+C_2^2)\Lambda=0.
\label{eqe8}
\end{equation}
Solving equation (\ref{eqe8}), we obtain the hyperbolic solution
\begin{equation}
\begin{split}
\Lambda(\xi)=& C_3\sqrt{g(\xi)}\sinh{\left(\int{3\sqrt{1+C_1^2+2C_1C_2\xi+\xi^2+C_2^2\xi^2}\mbox{ }(1+C_2^2)^2 \over 2\sqrt{1+C_2^2}\mbox{ }g(\xi)} d\xi\right)}\\ & + C_4\sqrt{g(\xi)}\cosh{\left(\int{3\sqrt{1+C_1^2+2C_1C_2\xi+\xi^2+C_2^2\xi^2}\mbox{ }(1+C_2^2)^2 \over 2\sqrt{1+C_2^2}\mbox{ }g(\xi)} d\xi\right)},
\end{split}
\label{eqe999}
\end{equation}
where we have defined 
\begin{equation}
g(\xi)=3C_1^2C_2^2-C_2^2-C_1^2-1+6C_1C_2\xi+6C_2^3C_1\xi+6C_2^2\xi^2+3C_2^4\xi^2+3\xi^2.
\end{equation}

For subalgebra $L_2$, we obtain the following two solutions. First, a trivial linear solution in $y$ for both $\varphi$ and $\psi$, and second a solution which is linear for $\varphi$ ($\varphi(y)=\pm iy+C_2$) but where $\psi$ is an arbitrary function of $y$. In particular, this includes the case where $\psi$ is a solitonic, bump, kink or multiple wave solution. From subalgebra $L_3$, we obtain solutions similar to those for $L_2$ except that the argument $y$ is replaced by $x$.

For subalgebra $L_4$, we should distinguish two separate cases. In the case where $\varepsilon=-1$ and $m=\pm 1$, we obtain the linear travelling wave
\begin{equation}
\varphi(x,y)=C_1(y\pm x)+C_2,\qquad \psi(x,y)=\underline{K_1}(y\pm x)+\underline{K_2},
\label{eqe11}
\end{equation}
where $C_1$ and $C_2$ are bosonic constants and $\underline{K_1}$ and $\underline{K_2}$ are fermionic constants.
For all other cases of $\varepsilon$ and $m$, we obtain two separate families of solutions, one of which is linear
\begin{equation}
\varphi(x,y)=C_1(y-mx)+C_2,\qquad \psi(x,y)=\underline{K_1}(y-mx)+\underline{K_2},
\label{eqe12}
\end{equation}
and the other, where $\varphi(x,y)$ is given by the specific expression
\begin{equation}
\varphi(x,y)=\pm\left(\varepsilon{m^2+1\over (m^2+\varepsilon)^2}\right)^{1/2}(y-mx)+C_2,
\label{eqe13}
\end{equation}
while the fermionic field $\psi$ is an arbitrary function of the single quantity $y-mx$. This allows us to consider a wide range of interesting physical phenomena, including travelling waves, center waves, bumps, kinks and multiple waves. In particular, we can choose fermionic solitary wave solutions, which are particularly interesting since they have been extensively studied and could give us information about the fermionic medium under consideration.

From subalgebras $L_6$ and $L_7$, we obtain linear solutions for $\varphi$ similar to those discussed above for $L_2$ and $L_3$ above, respectively. We still obtain solutions for the fermionic field $\psi$ which are arbitrary functions of $y$ and $x$ respectively. For subalgebra $L_8$, we get yet another linear solution similar to that of $L_4$, but also the specific solution
\begin{equation}
\varphi(x,y)={m\pm n\sqrt{\varepsilon(m^2+n^2)}\over m^2+\varepsilon n^2}\left(x-{m\over n}y\right)+C_2,\qquad \psi=\psi\left(x-{m\over n}y\right),
\label{eqe14}
\end{equation}
in terms of an arbitrary function for $\psi$. This represents a propagation wave in $x$ and $y$. Because of this freedom in $\psi$, we can consider, for example, bounded elementary, trigonometric, periodic and doubly periodic solutions, as well as Painlev\'{e} transcendents. Consequently, different physical phenomena can be considered.

\subsection{Non-splitting subalgebras and their reductions}

Let us turn now to the non-splitting subalgebras. Combining the two reduced equations for subalgebra ${\mathcal L}_{2}$, we obtain the following second order ordinary differential equation for $\varphi$ only
\begin{equation}
\left(1-\varepsilon(\varphi_y)^2\right)^2\varphi_{yy}-2\underline{\eta_1}\underline{\eta_2}(\varphi_y)^3+4\delta_{\varepsilon,1}\varphi_y\underline{\eta_1}\underline{\eta_2}=0,
\label{eqe15}
\end{equation}
where
\begin{equation}
\delta_{a,b}=\begin{cases} &1\mbox{, if }a=b\\ &0\mbox{, if }a\neq b\end{cases}
\end{equation}
is the usual Kronecker delta function. Equation (\ref{eqe15}) can be converted through the substitution $\omega=\varphi_y$ to a first order equation
\begin{equation}
\left(1-\varepsilon(\omega)^2\right)^2\omega_y-2\underline{\eta_1}\underline{\eta_2}\omega^3+4\delta_{\varepsilon,1}\omega\underline{\eta_1}\underline{\eta_2}=0.
\label{eqe16}
\end{equation}
In the case where $\varepsilon=-1$, equation (\ref{eqe16}) can be integrated to give the transcendental relation for $\omega$
\begin{equation}
4\omega^2\ln{\omega}-4\underline{\eta_1}\underline{\eta_2}\omega^2y+\omega^4-1-4C_1\underline{\eta_1}\underline{\eta_2}\omega^2=0.
\label{eqe17}
\end{equation}

The subalgebra ${\mathcal L}_{3}$ leads us to the quadratic solution
\begin{equation}
\varphi(x)=C_1x+C_2,\qquad \psi(x)=-{1\over 2(1-\varepsilon C_1^2)}\underline{\eta_2}x^2+\underline{K_1}x+\underline{K_2}+\underline{\eta_1}xy+\frac{1}{2}\underline{\eta_2}y^2.
\label{eqe18}
\end{equation}
In addition, in the case where $\underline{\eta_2}=0$, we also have
\begin{equation}
\varphi(x)=\pm \sqrt{\varepsilon}x+C_2,\qquad \psi(x)=\Lambda(x)+\underline{\eta_1}xy,
\label{eqe19}
\end{equation}
where $\Lambda$ is an arbitrary function of $x$. In similarity with the previous cases, we can adjust $\Lambda$ to various solitonic, compact suport functions.

The two reduced equations for subalgebra ${\mathcal L}_{4}$ can be combined into the following second order ordinary differential equation for $\varphi$
\begin{equation}
\begin{split}
&\varepsilon(1+m^2)^2\varphi_{\xi\xi}-2(1+m^2)(1+\varepsilon m^2)^2(\varphi_{\xi})^2\varphi_{\xi\xi}+\varepsilon(1+\varepsilon m^2)^4(\varphi_{\xi})^4\varphi_{\xi\xi}\\ &+2(1-m^2)\underline{\eta_1}\underline{\eta_2}(\varphi_{\xi})^3-4\delta_{\varepsilon,1}\varphi_{\xi}\underline{\eta_1}\underline{\eta_2}=0,
\end{split}
\label{eqe20}
\end{equation}
which can be reduced to the first order equation for $\omega$
\begin{equation}
\begin{split}
&\varepsilon(1+m^2)^2\omega_{\xi}-2(1+m^2)(1+\varepsilon m^2)^2\omega^2\omega_{\xi}+\varepsilon(1+\varepsilon m^2)^4\omega^4\omega_{\xi}+2(1-m^2)\underline{\eta_1}\underline{\eta_2}\omega^3\\ &-4\delta_{\varepsilon,1}\omega\underline{\eta_1}\underline{\eta_2}=0.
\end{split}
\label{eqe21}
\end{equation}
This equation does not have the Painlev\'{e} property, implying that there are additional singularities besides fixed poles.

For the case of subalgebra ${\mathcal L}_{6}$, the reduced equations from table 3 combine to give the following equation for $F$
\begin{equation}
\left(1-\varepsilon(F_y)^2\right)^2F_{yy}+\varepsilon{2\over m^2}\underline{\eta_1}\underline{\eta_2}(F_y)^3+{2\over m^4}(1-2m^2\delta_{\varepsilon,1})\underline{\eta_1}\underline{\eta_2}F_y=0.
\label{eqe22}
\end{equation}
This equation can be transformed to the first order ordinary differential equation for $\omega$
\begin{equation}
\left(1-\varepsilon\omega^2\right)^2\omega_{y}+\varepsilon{2\over m^2}\underline{\eta_1}\underline{\eta_2}\omega^3+{2\over m^4}(1-2m^2\delta_{\varepsilon,1})\underline{\eta_1}\underline{\eta_2}\omega=0.
\label{eqe23}
\end{equation}

For the case of subalgebra ${\mathcal L}_{7}$, combination of the reduced equations gives us the following equation for $F$
\begin{equation}
\left(1-\varepsilon(F_x)^2\right)^2F_{xx}+{2\over m^5}\underline{\eta_1}\underline{\eta_2}\left(m^2(F_x)^2+\varepsilon-2m^2\delta_{\varepsilon,1}\right)=0.
\label{eqe24}
\end{equation}
This equation corresponds to the first order ordinary differential equation
\begin{equation}
\left(1-\varepsilon\omega^2\right)^2\omega_{x}+{2\over m^5}\underline{\eta_1}\underline{\eta_2}\left(m^2\omega^2+\varepsilon-2m^2\delta_{\varepsilon,1}\right)=0.
\label{eqe25}
\end{equation}

For subalgebra ${\mathcal L}_{8}$, the two reduced equations give us the equation
\begin{equation}
\begin{split}
& \Bigg{(}(m^2+\varepsilon n^2)^4(F_{\xi})^4-4\varepsilon m(n^2+\varepsilon m^2)^3(F_{\xi})^3\\ & +\Big{(}6m^2(m^2+\varepsilon n^2)^2-2\varepsilon n^2(n^4+\varepsilon m^4)(n^2+\varepsilon m^2)-4m^2n^4(n^2+m^2)\delta_{\varepsilon,1}\Big{)}(F_{\xi})^2\\ & +4m\Big{(}n^2(n^4+\varepsilon m^4)-\varepsilon m^2(n^2+\varepsilon m^2)+2m^2n^4\delta_{\varepsilon,1}\Big{)}F_{\xi}\\ & +\Big{(}m^2(1-\varepsilon n^2)-\varepsilon n^4 \Big{)}^2\Bigg{)}\mbox{ }F_{\xi\xi}\\ & + \Bigg{(}-2\varepsilon mn^3(m^2+\varepsilon n^2)(F_{\xi})^3+2n^3(n^2+3\varepsilon m^2)(F_{\xi})^2\\ & +(4mn^5\delta_{\varepsilon,1}-6\varepsilon mn^3)F_{\xi}+2\varepsilon n^3-4n^5\delta_{\varepsilon,1}\Bigg{)}\mbox{ }\underline{\eta_1}\underline{\eta_2}=0.
\end{split}
\label{megaeq}
\end{equation}
Through the substitution $\omega=\varphi_y$, equation (\ref{megaeq}) can be reduced to a first order equation.

\section{Conclusions}

In this paper, we have formulated supersymmetric generalizations for the Gaussian irrotational fluid flow equation (\ref{qwertynot}) and its modified version (\ref{qwertynot2}). It is interesting and significant to note that the symmetry superalgebras for our supersymmetrized Gaussian fluid model equation and for its modified version were isomorphic to each other, as well as to that of the supersymmetric Born-Infeld scalar equation (\ref{qwerty1}) as investigated in \cite{Hariton}. This is in contrast to the classical case, where the symmetries for equation (\ref{qwertynot}) included a rotation which was not present in (\ref{qwertynot2}). The three translations in $x$, $y$ and $\phi$ present in the classical case are also present in the supersymmetric case, while the dilation in $x$, $y$ and $\phi$ is modified to a dilation which also includes the fermionic field $\psi$. However, the rotation transformation present in the classical fluid equation (\ref{qwertynot}) is not preserved in the corresponding supersymmetric equations (\ref{eqb7}) and (\ref{eqb8}) for $\varepsilon=1$. Conversely, a new translation in $\psi$ and two more fermionic transformations (for specific values of the parameters $a$, $b$, $c$ and $d$) are present in the supersymmetric case.

Through the use of the symmetry reduction method, invariant solutions of the classical and supersymmetric Gaussian fluid were systematically constructed. Solutions of the classical Gaussian flow equation (\ref{qwertynot}) include a kink-type fluid density (\ref{kinksolution}) representing a condensation wave and a solution (\ref{lambertsolution}) expressed in terms of the Lambert function. The kink-type expression (\ref{kinksolution}) is also present as a solution of the supersymmetric equations (\ref{eqc6}) and (\ref{eqc7}) with $\varepsilon=1$, where it represents the bosonic part of the solution (\ref{eqe3biss}) and (\ref{eqe6biss}). Its fermionic component is given in terms of radicals and trigonometric functions. The classical modified equation (\ref{qwertynot2}) possesses a solution (\ref{eqebis3ZZ}) expressed in terms of elliptic functions, whose imaginary part can be used to approximate a bump function. This solution also appears as the bosonic component of a solution (\ref{eqe3}) and (\ref{eqe6}) of the supersymmetric equations (\ref{eqc6}) and (\ref{eqc7}) with $\varepsilon=-1$. For the invariant solutions of both supersymmetric extensions, there are instances (eg. (\ref{eqe13}), (\ref{eqe14}) and (\ref{eqe19})) where the fermionic field $\psi$ is an arbitrary function of a single quantity. This permits us consider a wide range of possibilities for the fermionic medium. Equations (\ref{eqb7}) and (\ref{eqb8}) represent the static case which involves two spatial variables. It is expected that a supersymmetric extension of the Gaussian fluid flow equations in $(2+1)$ dimensions may yield richer classes of physically relevant solutions. 

{\bf ACKNOWLEDGMENTS}

The authors would like to thank Professor R. Jackiw of MIT for helpful and interesting discussions on the topic of this paper. This work is supported in part by funds provided by the U.S. Department of Energy (D.O.E.) under cooperative research agreement DEFG02-05ER41360, and by research grants from NSERC of Canada and FQRNT of Qu\'{e}bec. A.J.H. acknowledges support by the FQRNT of Canada under their postdoctoral research fellowship program. 


{}

\end{document}